\documentclass[%
reprint,
superscriptaddress,
amsmath,amssymb,
aps,
]{revtex4-1}

\usepackage{booktabs}
\usepackage{multirow}

\usepackage{subcaption}

\usepackage{bm} 
\usepackage{graphicx}
\usepackage{dcolumn}
\usepackage{bm}
\usepackage{hyperref}

\begin{document}

	\preprint{APS/123-QED}
	
	\title{Yang-Mills field modified RN black hole and the Strong Cosmic Censorship Conjecture}
	
	\author{Zhiqin Tu}
	
	\author{Meirong Tang}

	\author{Zhaoyi Xu}%
	\email{zyxu@gzu.edu.cn(Corresponding author)}
	\affiliation{%
		College of Physics,Guizhou University,Guiyang,550025,China
	}%

	\begin{abstract}
To address the singularity problem in black hole physics, Penrose proposed the cosmic censorship conjecture . However, whether this conjecture holds in different types of black holes, especially for the Strong Cosmic Censorship Conjecture (SCCC), remains an issue worth further exploration. This study investigates the Yang-Mills field modified RN black hole, which provides new possibilities for exploring the Strong Cosmic Censorship Conjecture (SCCC) due to its unique nonlinear dynamical properties. In this work, the perturbation equations of two scalar fields for the Yang-Mills field modified RN black hole are derived. By combining the WKB method and the Prony method, the quasi-normal modes frequencies of the perturbations are analyzed. Additionally, constraints on the charge-to-mass ratio of the charged and massive scalar field are obtained using the Weak Gravity Conjecture (WGC), and it is shown that the Yang-Mills field modified RN black hole satisfies the SCCC conditions in the extreme state. The study reveals that the Yang-Mills field charge $q_{YM}$ significantly affects the damping effect and gravitational blueshift effect of the perturbations, causing the RN black hole in the extreme state to transition from violating the SCCC to satisfying it. Furthermore, when the charge-to-mass ratio $q/m$ of the charged and massive scalar field satisfies the WGC condition, the scalar field perturbations destabilize the Cauchy horizon of the Yang-Mills field modified RN black hole in the extreme state, thus confirming the applicability of the SCCC under extreme conditions. This study highlights the key role of nonlinear fields in singularity theory and gravitational dynamics, providing important support for the completeness of general relativity and offering new insights into the theoretical research of black holes in extreme states.

\end{abstract}

\maketitle


\section{\label{sec:level1}Introduction}

General relativity provides a theoretical framework for describing gravitational phenomena. Its core lies in the Einstein field equations, which reveal how mass, energy, and momentum shape the geometry of spacetime. The curvature of spacetime determines the motion of objects, while the distribution of mass and energy, in turn, influences the curvature of spacetime. Black holes, as one of the key predictions of general relativity\cite{Berti:2015itd}\cite{Shipley:2016omi}, are among the most extreme objects in the universe. Since Einstein proposed the field equations, many classical black hole solutions have emerged, such as the Schwarzschild black hole\cite{Ovalle:2024wtv}\cite{Doran:2006dq}, the Reissner-Nordström black hole\cite{Cano:2019ycn}\cite{Wang:2021woy}, and the Kerr black hole\cite{Teukolsky:2014vca}\cite{Gralla:2019drh}. These black hole solutions demonstrate that the external spacetime geometry of a black hole is entirely determined by its mass, angular momentum, and charge, reflecting the core idea of the no-hair theorem\cite{Israel:1967wq}\cite{Carter:1971zc}: black holes have no additional observable structures, and all their characteristics are determined by just these three physical quantities—mass, angular momentum, and charge. However, inside a black hole exists a region where the curvature of spacetime becomes infinite, known as a singularity\cite{Senovilla:2014gza}\cite{Kunzinger:2021das}. The formation of a singularity signifies that the structure of spacetime reaches extreme distortion, rendering general relativity and classical physical theories incapable of making effective predictions. If the singularity is not covered by the black hole's event horizon, it would lead to the formation of a naked singularity, which would destroy the causal structure of spacetime and the determinism of physical laws, ultimately posing a serious challenge to the applicability and completeness of general relativity.

In order to maintain the applicability and completeness of general relativity and to address the potential issues posed by singularities, Roger Penrose proposed the famous cosmic censorship conjecture in the 1960s. This conjecture aims to limit the influence of singularities on the external universe, thereby preserving the causal structure and predictive power of physical laws. The Cosmic Censorship Conjecture can be divided into two main forms: the Weak Cosmic Censorship Conjecture (WCCC)\cite{Penrose:1969pc} and the Strong Cosmic Censorship Conjecture (SCCC)\cite{Penrose:1980ge}. The WCCC requires that singularities must be completely surrounded by the event horizon of a black hole, making it impossible for external observers to directly detect the singularity. This assumption provides a theoretical constraint on the causal relationships of spacetime in general relativity, meaning that naked singularities will not form, thus preventing the breakdown of physical laws due to naked singularities. In contrast, the SCCC is a further strengthening of the WCCC, requiring that the causal structure inside a black hole should not extend continuously to the external world, avoiding the appearance of unpredictable regions. Even within the event horizon of a black hole, spacetime should avoid causality violations to ensure the predictability and global stability of spacetime solutions. In other words, it demands that the Cauchy horizon inside a black hole must be unstable, thus preventing the collapse of the causal structure of spacetime. In classical black hole models, the WCCC has been extensively validated; for example, in Schwarzschild and Reissner-Nordstr\"om black holes, singularities are generally covered by the event horizon\cite{Zhao:2024qzg}\cite{Zhao:2024lts}\cite{Tang:2023sig}\cite{Shaymatov:2020wtj}\cite{Wang:2019bml}. But, some theoretical or specific perturbative scenarios suggest that the WCCC may be violated, where singularities could become naked and not be obscured by an event horizon\cite{Meng:2024els}\cite{Zhao:2023vxq}\cite{Lin:2024deg}\cite{Miyamoto:2011tt}. However, the validity of the SCCC depends heavily on the stability of the Cauchy horizon. The SCCC typically holds in asymptotically flat spacetimes\cite{Destounis:2021lum}\cite{Caspar:2012ux}\cite{Cao:2024kht}\cite{Yi-Wen:2024ali}, especially for Kerr and Reissner-Nordström black holes, because the exponential blueshift effect caused by perturbations makes the Cauchy horizon unstable and non-extensible. However, in asymptotically de Sitter spacetimes, the positive cosmological constant causes competition between the exponential decay of external perturbations and the gravitational blueshift effect, which may lead to the failure of the SCCC. In particular, for near-extremal charged Reissner-Nordström-de Sitter black holes, the SCCC is violated by perturbations from scalar fields, fermionic fields, and gravitational-electromagnetic fields\cite{Tu:2024huh}\cite{Jiang:2023bpr}\cite{Shao:2023qlt}\cite{Dias:2019ery}\cite{Casals:2020uxa}\cite{Destounis:2019omd}\cite{Davey:2024xvd}\cite{Dias:2020ncd}.

This paper studies the Reissner-Nordström (RN) black hole model modified by Yang-Mills fields\cite{Gomez:2023qyv}, demonstrating its unique spacetime geometric properties by introducing nonlinear corrections from the non-Abelian Yang-Mills fields. This provides a new perspective for testing the SCCC. Compared to the traditional RN black hole, this nonlinear correction not only changes the relative positions of the black hole's event horizon and Cauchy horizon, increasing the distance between them, but also significantly enhances the blueshift effect of the black hole, thereby exacerbating the instability of the Cauchy horizon. Through the analysis of the propagation characteristics of scalar fields in this modified spacetime, especially the perturbative behavior of neutral massless scalar fields and charged massive scalar fields on the black hole, this paper reveals new dynamical processes in this spacetime background and provides a theoretical validation of the applicability of the SCCC in complex spacetimes. The introduction of the Yang-Mills field also has a significant impact on the black hole's accretion dynamics, increasing the accretion efficiency and altering the radius of the innermost stable circular orbit (ISCO)\cite{Gomez:2023qyv}, providing new insights for understanding high-accretion-rate black hole environments and astrophysical observations. In conclusion, this paper aims to explore whether the Yang-Mills field modified RN black hole obeys the SCCC under scalar field perturbations and whether this model can maintain the completeness and applicability of general relativity.

The structure of this paper is arranged as follows: In Section \ref{sec:level2} briefly introduces the theoretical background of the Yang-Mills field modified RN black hole. In Section \ref{sec:level3} explains the perturbation equation under scalar field perturbations and the conditions for the validity of the SCCC. In Section \ref{sec:level4}  briefly describes the research methods adopted in this study. In Section \ref{sec:level5} presents the main research results, analyzes the behavior of the Yang-Mills field modified RN black hole under two types of scalar field perturbations, and explores the significance of these results in verifying the SCCC. In Section \ref{sec:level6} presents the conclusions and discussions.

\section{\label{sec:level2}Yang-Mills field modified RN black hole}
	
The Yang-Mills field modified RN black hole studied in this paper is unique in that it not only includes parameters for mass and electromagnetic charge but also introduces a modified charge induced by the nonlinear Yang-Mills field. This black hole is an extension of the standard RN black hole through the incorporation of a nonlinear field, which significantly affects the black hole's geometric structure and physical properties. For example, the introduction of the corrected charge changes the event horizon structure and the location of stable circular orbits, enhances the accretion efficiency, and avoids the appearance of naked singularities\cite{Gomez:2023qyv} . This black hole model provides a new perspective for studying the interaction between non-Abelian gauge fields and gravity in general relativity, and it also helps to explore the applicability of the SCCC in more complex scenarios.
	
To derive the solution for this black hole, the following steps can be taken:
\begin{equation}
	S = \frac{1}{2} \int \sqrt{-g} \, d^4 x \, \left[ R - F_M - F_p^{YM} \right].
		\label{ep1}
\end{equation}
Here, \( R \) is the Ricci scalar, \( F_M = F_{\mu\nu} F^{\mu\nu} \) is the Maxwell field invariant, and \( F_p^{YM} = \text{Tr} \left( F_{\lambda\delta}^{(a)} F^{\lambda\delta (a)} \right)^p \) is the Yang-Mills field invariant, with the nonlinearity described by the parameter \( p \). By varying the action and taking the derivative with respect to the metric tensor \( g_{\mu\nu} \), the modified Einstein field equations can be obtained:
\begin{equation}
G_{\mu\nu} + \Lambda g_{\mu\nu} = T_{\mu\nu}.
	\label{ep2}
\end{equation}
Here, the energy-momentum tensor \( T_{\mu\nu} \) includes contributions from the Maxwell field and the nonlinear Yang-Mills field. Now, considering a spherically symmetric spacetime in the Schwarzschild coordinate system, its metric can be written as\cite{Gomez:2023qyv}:
\begin{equation}
ds^2 = -f(r) dt^2 + f(r)^{-1} dr^2 + r^2 \left( d\theta^2 + \sin^2\theta \, d\phi^2 \right).
	\label{ep3}
\end{equation}
The field equation simplifies to an ordinary differential equation:
\begin{equation}
	\frac{d}{dr} \left( r f(r) \right) = 1 - \frac{Q^2}{r^2} - \frac{2^p q_{YM}^{2p}}{r^{4p - 2}}.
	\label{ep4}
\end{equation}
where \( Q \) is the electromagnetic charge and \( q_{YM} \) is the Yang-Mills field charge. By integrating this equation and choosing an appropriate integration constant, the analytical expression for the corrected black hole can be obtained as\cite{Gomez:2023qyv}:
\begin{equation}
	f(r) = 1 - \frac{2M}{r} + \frac{Q^2}{r^2} + \frac{Q_{YM}}{r^{4p - 2}}.
	\label{eq5}
\end{equation}
Here \( Q_{YM} = \frac{2^{2p - 1}}{4p - 3} q_{YM}^{2p} \)\cite{Mazharimousavi:2009mb}\cite{HabibMazharimousavi:2008zz}. The metric function of this black hole solution contains three main contributions: \( 1 - \frac{2M}{r} \) represents the mass term of the Schwarzschild black hole, \( \frac{Q^2}{r^2} \) is the charge term of the RN black hole, and \( \frac{Q_{YM}}{r^{4p - 2}} \) is the modified term introduced by the nonlinear Yang-Mills field. The study shows that the nonlinear Yang-Mills field has a profound impact on the geometry of black holes. It not only alters the position of the event horizon and the radius of stable circular orbits but also, under certain conditions, enhances accretion efficiency and effectively prevents the formation of naked singularities, thereby preserving the physical integrity of black holes. Particularly in the case \( p = \frac{1}{2} \), this formula not only possesses analytical simplicity but also demonstrates significant physical meaning, providing a new perspective for studying the interaction between non-Abelian gauge fields and gravity.

When \( q_{YM} = 0 \), the Yang-Mills field modified RN black hole degenerates into the standard RN black hole. Therefore, in equation (\ref{eq5}), the parameter \( p \) must be greater than \( 0 \). If \( p < 0 \), the Yang-Mills field modified RN black hole cannot reduce to the RN black hole, and when \( q_{YM} = 0 \), it will produce infinite values, which contradict classical physics expectations.In the standard RN black hole (\( f(r) = 1 - \frac{2M}{r} + \frac{Q^2}{r^2} \)), when the mass \( M \) and charge \( Q \) satisfy \( M = Q \), the RN black hole will be in an extreme state, where the SCCC is often violated. This paper primarily investigates whether the modified term introduced by the Yang-Mills field (\( \frac{Q_{YM}}{r^{4p - 2}} \)) helps the Yang-Mills field modified RN black hole obey the SCCC under conditions \( M = 1 \) and \( Q = 1 \).For the Yang-Mills field modified RN black hole with mass \( M = 1 \) and charge \( Q = 1 \), to ensure it has two positive real roots, equation (\ref{eq5}) yields:\( r^{4p-4} (r - 1)^2 = -\frac{2^{2p - 1}}{4p - 3} q_{YM}^{2p} \)
Thus, to ensure the physical validity of the black hole, the parameter \( p \) must be less than \( \frac{3}{4} \).In summary, we obtain the range of the parameter \( p \): \( 0 < p < \frac{3}{4} \).

\begin{figure*}
		\centering
		\resizebox{\linewidth}{!}{\includegraphics*{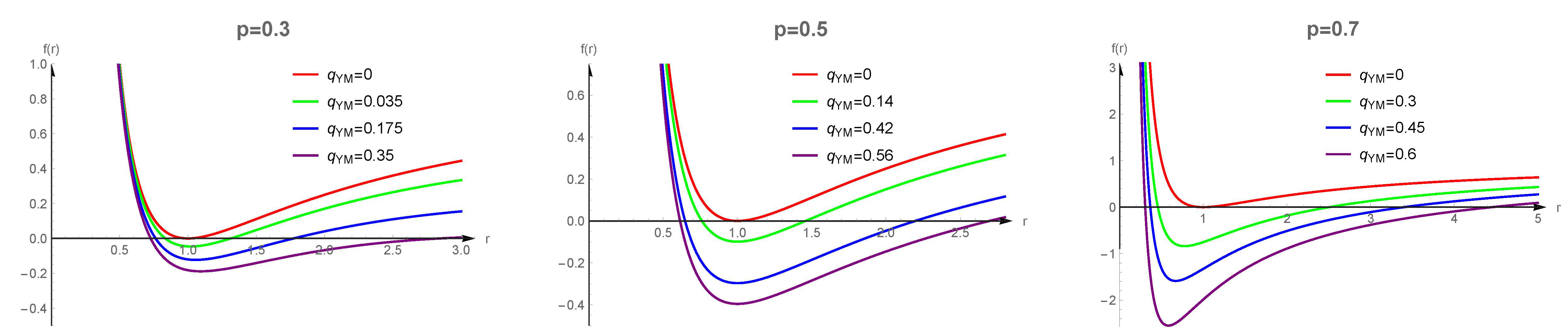 }}
		\captionsetup{justification=raggedright, singlelinecheck=false}  
		\caption{Under the assumptions of black hole mass $M = 1$ and electromagnetic charge $Q = 1$, the effects of the Yang-Mills field modified on the event horizon radius of the RN black hole are analyzed. The variation of the event horizon radius is demonstrated for different values of the parameter $p$ and the Yang-Mills field charge $q_{YM}$.
		}
		\label{fig1}
\end{figure*}
	
In Figure \ref{fig1}, we show how the horizon radius of the Yang-Mills field modified RN black hole changes with the Yang-Mills field charge \( q_{YM} \) under different values of the parameter \( p \), where the mass \( M = 1 \) and the electromagnetic charge \( Q = 1 \). By observing Figure \ref{fig1}, we can see that when the Yang-Mills field charge \( q_{YM} = 0 \), the black hole degenerates into the standard RN black hole, where the black hole is in an extreme state with only one horizon existing. When the Yang-Mills field charge \( q_{YM} > 0 \), the Yang-Mills field modified RN black hole has two horizons: the event horizon \( r_{h_+} \) and the Cauchy horizon \( r_{h_-} \). However, as the Yang-Mills field charge \( q_{YM} \) gradually increases, not only do both the event horizon and the Cauchy horizon exist simultaneously, but the difference between the two horizon radii becomes increasingly larger. This phenomenon not only highlights the effect of the Yang-Mills field on the RN black hole structure but also favors the Yang-Mills field modified RN black hole in obeying the SCCC, thereby maintaining the validity of general relativity.

\section{\label{sec:level3} Scalar Field Perturbation and Strong Cosmic Censorship Conjecture}

The Yang-Mills field modified RN black hole is an important model in nonlinear gravitational theory, and its complex horizon structure provides a new perspective for studying the SCCC. In this modified model, scalar field perturbations are key to analyzing the stability of the Cauchy horizon and its dynamical behavior. By deriving the perturbation equation of the scalar field in this spacetime background, one can analyze the effects of gravitational blueshift and decay on the stability of the Cauchy horizon. Additionally, it is possible to derive different types of black holes from the quasi-normal modes(QNMs) analysis of scalar field perturbations\cite{Liu:2021xfb}. In summary, this unique nonlinear corrected black hole model provides a possibility to test the validity of the SCCC.
	
\subsection{\label{sec:level3.1}Scalar field perturbation}
In this subsection, we study the evolution of scalar field perturbations in the spacetime background of the Yang-Mills field modified RN black hole. For a neutral and massless scalar field\cite{Cardoso:2017soq}, the Klein-Gordon equation can be written as:
\begin{equation}
		\frac{1}{\sqrt{-g}} \partial_\mu \left( g^{\mu \nu} \sqrt{-g} \, \partial_\nu \Phi \right) = 0.
		\label{eq6}
\end{equation}
For a charged and massive scalar field\cite{Konoplya:2013rxa}, the Klein-Gordon equation can be written as:
\begin{equation}
	\begin{aligned}
		\frac{1}{\sqrt{-g}} \partial_\mu \left( g^{\mu \nu} \sqrt{-g} \partial_\nu \Phi \right) 
		- 2 i q g^{\mu \nu} A_\mu \partial_\nu \Phi \\
		- q^2 g^{\mu \nu} A_\mu A_\nu \Phi - m^2 \Phi = 0.
	\end{aligned}
	\label{eq7}
\end{equation}
In four-dimensional spacetime, the mass and charge of the scalar field are represented by the parameters $m$ and $q$, respectively. The electromagnetic potential vector of the black hole can be expressed as $A_\mu = (-Q/r, 0, 0, 0)$, where $Q$ is the charge of the black hole and $r$ represents the radial distance. This expression describes the electrostatic potential distribution generated by the black hole at the radial position $r$. Due to the spherical symmetry of spacetime, the scalar field can be expanded in terms of spherical harmonics.
\begin{equation}
	\Phi(t, r, \theta, \phi) = \sum_{m' \ell} \frac{\varphi(r)}{r} Y_{\ell m'}(\theta, \phi) e^{-i \omega t}.
	\label{eq8}
\end{equation}
The frequency of the scalar field \( \Phi \) is denoted by \( \omega \), and \( \varphi(r) \) is its corresponding radial wave function. The angular momentum quantum number \( l \) only takes non-negative integers, while the magnetic quantum number \( m' \) is set to zero in this paper. The spherical harmonics \( Y_{lm}(\theta, \phi) \) describe the angular distribution characteristics. We focus only on the perturbed radial equation, which can be expressed as:

\begin{equation}
	\left( \frac{d^2}{dr_*^2} + \omega^2 - V(r) \right) \varphi(r) = 0.
	\label{eq9}
\end{equation}
The tortoise coordinate \( dr_* \) is defined by the differential equation \( dr_* = \frac{dr}{f(r)} \), with its range extending from negative infinity at the horizon to positive infinity in space. Its expression can be further given as\cite{Sarkar:2023rhp}:

\begin{equation}
	r_* = r - \frac{1}{2k_{h_-}} \ln\left( \frac{r}{r_{h_-}} - 1 \right) + \frac{1}{2k_{h_+}} \ln\left( \frac{r}{r_{h_+}} - 1 \right).
	\label{eq10}
\end{equation}
Here, \( k_{h_+} \) and \( k_{h_-} \) represent the surface gravities at the event horizon and Cauchy horizon, respectively. Their expressions are given by:
\begin{equation}
k_{h_+} = \left| \frac{1}{2} f'(r_{h_+}) \right|, \quad k_{h_-} = \left| \frac{1}{2} f'(r_{h_-}) \right|.
	\label{eq11}
\end{equation}
The effective potential for a neutral and massless scalar field is given by:
\begin{equation}
V_{\text{eff-1}}(r) = f(r) \left[ \frac{\ell(\ell + 1)}{r^2} + \frac{f'(r)}{r} \right].
	\label{eq12}
\end{equation}
The effective potential for a charged and massive scalar field becomes:
\begin{align}
	V_{\text{eff-2}}(r) =& \, \omega^2 - f(r) \left( \frac{\ell(\ell+1)}{r^2} + \frac{f'(r)}{r} \right) \notag \\
	& + \frac{q^2 Q^2}{r^2} - \frac{2\omega q Q}{r} - m^2 f(r).
	\label{eq13}
\end{align}
Due to the absorptive nature of the black hole event horizon, all perturbations reaching the horizon are completely absorbed without any reflection. At the event horizon, the perturbations take the form of purely incoming waves, while at infinity, they appear as purely outgoing waves. For a neutral and massless scalar field, the boundary conditions are as follows:
\begin{equation}
	\varphi(r) \sim 
	\begin{cases}
		e^{-i \omega  r_*}, & r_* \to -\infty, \\
		e^{i \omega r_*}, & r_* \to +\infty.
	\end{cases}
	\label{eq14}
\end{equation}
The boundary conditions for a charged and massive scalar field are:
\begin{equation}
	\varphi(r) \sim 
	\begin{cases}
		e^{-i \left( \omega - \frac{qQ}{r_{h_+}} \right) r_*}, & r_* \to -\infty, \\
		e^{i \omega r_*}, & r_* \to +\infty.
	\end{cases}
	\label{eq15}
\end{equation}
When the effective potential gradually approaches zero at infinity, the perturbations in this region decay and eventually vanish as the distance increases. Utilizing this property, appropriate boundary conditions can be set, and by solving the perturbation equation, the discrete QNMs frequencies can be obtained. These frequencies reflect the inherent oscillation modes of the system.

\subsection{\label{sec:level3.2}Strong Cosmic Censorship Conjecture}
In black hole physics, the Strong Cosmic Censorship Conjecture (SCCC)\cite{Penrose:1980ge} is a significant theoretical problem in general relativity. The core claim of the SCCC is that classical solutions of general relativity should have a well-behaved global causal structure and determinism to ensure the global predictability of physical laws. The instability of the Cauchy horizon is key to ensuring that the classical solutions cannot be smoothly extended, and it is also an important criterion for verifying whether the SCCC holds. If the Cauchy horizon is stable, the causal structure inside the black hole would remain smooth and stable, which contradicts the requirements of the SCCC, leading to the conjecture's failure. Therefore, studying the dynamic evolution and decay characteristics of black holes under perturbations is crucial for verifying the SCCC.

In the mathematical description of black hole perturbations, the commonly used formula to describe the decay of perturbations over time is\cite{Singha:2022bvr}:
\begin{equation}
	\varphi \sim \exp(-Im\omega u)\varphi_0.
	\label{eq16}
\end{equation}
Here,  \(  Im\omega \) represents the imaginary part of the QNMs frequency, which determines the decay rate of black hole perturbations. The larger the imaginary part, the faster the decay. The parameter \( u \) is defined as \( u = t - r_* \), used to describe the temporal evolution of the perturbations, where \( t \) is the time coordinate, and \( r_* \) is referred to as the "tortoise coordinate." Furthermore, under the blueshift effect caused by the strong gravitational field, the change in the scalar field intensity near the horizon can be expressed as\cite{Singha:2022bvr}:
\begin{equation}
	|\varphi_{r_h}|^2 \sim \exp(k_i u)|\varphi_0|^2.
	\label{eq17}
\end{equation}
The surface gravity \( \kappa_i \) is a measure of the gravitational acceleration at the horizon, reflecting the intensity of the gravitational field around the horizon. It can be calculated using the formula \( \kappa_i = \left| \frac{1}{2} f'(r_i) \right| \), where \( f(r) \) is the metric function, and the derivative at \( r = r_i \) determines the specific value of the surface gravity. Higher surface gravity indicates a stronger gravitational field in the horizon region. In the study of the SCCC, \( \kappa_i \) and \( \text{Im} \, \omega \) are considered key analytical indicators. Whether the SCCC holds depends on which effect dominates, and their ratio can be expressed as:
\begin{equation}
	\beta = -\frac{\text{Im}\, \omega}{k_i}
	\label{eq18}.
\end{equation}
When \( \beta < \frac{1}{2} \), the decay effect of the scalar field is weaker than the gravitational blueshift effect, causing the Cauchy horizon of the black hole to become unstable or even break down, thereby satisfying the SCCC. However, when \( \beta > \frac{1}{2} \), the decay effect of the scalar field is stronger than the gravitational blueshift effect, and the Cauchy horizon of the black hole remains stable, leading to a violation of the SCCC. Therefore, the SCCC holds when \( \beta < \frac{1}{2} \), but it may fail when \( \beta > \frac{1}{2} \).

\section{\label{sec:level4}METHOD INTRODUCTION}

In this chapter, we will provide a detailed introduction to the main analytical and numerical methods used in this study. The aim is to analyze the QNMs frequencies of the Yang-Mills field modified RN black hole under scalar field perturbations and to verify their impact on the stability of the Cauchy horizon. To achieve this, we combine multiple methods to systematically study the dynamical properties of the Yang-Mills field modified RN black hole from different perspectives. First, the WKB method\cite{Wentzel:1926aor}\cite{Kramers:1926njj} is employed to compute the QNMs frequencies of the Yang-Mills field modified RN black hole under neutral and massless scalar field perturbations, obtaining key numerical results that lay the foundation for subsequent research. Second, the Prony method\cite{Dubinsky:2024gwo}\cite{Berti:2007dg} is used to decompose and fit the time-domain signals, further analyzing the perturbation behavior of neutral and massless scalar fields on the Yang-Mills field modified RN black hole. This allows us to extract the corresponding QNMs frequencies and compare them with the WKB method results to ensure the reliability and validity of the data. In addition, for the extreme Yang-Mills field modified RN black hole under charged and massive scalar field perturbations, the WGC\cite{Arkani-Hamed:2006emk}\cite{Harlow:2022ich} is introduced to impose constraints on the conditions for the validity of the SCCC. The organic combination of these methods not only comprehensively reveals the behavior of the Yang-Mills field modified RN black hole under scalar field perturbations but also provides theoretical support and numerical evidence for verifying the conditions under which the SCCC holds. In the following subsections, we will explain the principles of each method in detail.

\subsection{\label{sec:level4.1} WKB method}
The WKB method (Wentzel-Kramers-Brillouin)\cite{Wentzel:1926aor}\cite{Kramers:1926njj} is an important tool in asymptotic analysis, originally introduced in the field of quantum mechanics to obtain asymptotic solutions to complex wave equations. With its subsequent generalization and development, its applications in black hole research have gradually emerged, particularly demonstrating significant advantages in solving complex potential differential equations. For high-frequency modes, the WKB method can yield extremely accurate results, and by introducing higher-order modified, it further enhances the reliability of numerical solutions. In analyzing the quasinormal mode frequencies (QNMs) of black hole gravitational wave signals, the WKB method has played a significant role, providing strong support for testing general relativity and serving as a groundbreaking theoretical tool for exploring new physical phenomena.

According to Equation (\ref{eq9}), the WKB method\cite{Iyer:1986np}\cite{Iyer:1986nq} is used to derive the frequencies of the QNMs.
\begin{equation}
	iK - \left(n + \frac{1}{2}\right)  - \Lambda(n) = \Omega(n).
	\label{eq19}
\end{equation}
which
\begin{equation}
	K = \frac{V_0}{\sqrt{2V_0^{(2)}}},
	\label{eq20}
\end{equation}

\begin{equation}
	\Lambda(n) = \frac{1}{\sqrt{2V_0^{(2)}}} \left[ \frac{(a^2 + \frac{1}{4})V_0^{(4)}}{8 V_0^{(2)} }- \frac{(60a^2 + 7)}{288} \left( \frac{V_0^{(3)}}{V_0^{(2)}} \right)^2 \right],
	\label{eq21}
\end{equation}

\begin{widetext} 
	\begin{align}
		\label{eq22}
		\Omega(n) = & \, \frac{n + \frac{1}{2}}{2V_0^{(2)}} \left[ \frac{5(188a^2 + 77)}{6912} \left( \frac{V_0^{(3)}}{V_0^{(2)}} \right)^4 - \frac{(100a^2 + 51)}{384} \frac{(V_0^{(3)})^2V_0^{(4)}}{(V_0^{(2)})^3} \right] \nonumber \\
		& + \frac{n + \frac{1}{2}}{2V_0^{(2)}} \left[ \frac{(68a^2 + 67)}{2304} \left( \frac{V_0^{(4)}}{V_0^{(2)}} \right)^2 + \frac{(28a^2 + 19)}{288} \frac{V_0^{(3)}V_0^{(5)}}{(V_0^{(2)})^2} - \frac{(4a^2 + 5)}{288} \frac{V_0^{(6)}}{V_0^{(2)}} \right],
	\end{align}
\end{widetext}
In the above formula, \( V_0^{(k)} = \left. \frac{d^k V}{dx^k} \right|_{r=r_0} \) represents the \( k \)-th derivative of the effective potential at \( r = r_0 \), and \( a = n + \frac{1}{2} \). Therefore, according to Equation (\ref{eq19}), we can obtain:
\begin{align}
	\omega^2 ={} & \left[ V_0 + \left( -2V_0^{(2)} \right)^{1/2} \tilde{\Lambda}(n) \right] \notag \\
	& - i \left( n + \frac{1}{2} \right) \left( -2V_0^{(2)} \right)^{1/2} \left[ 1 + \tilde{\Omega}(n) \right].
	\label{eq23}
\end{align}
In the above expression
\begin{equation}
	\tilde{\Lambda} = \frac{\Lambda}{i}, \quad \tilde{\Omega} = \frac{\Omega}{\left( n + \frac{1}{2} \right)}.
	\label{eq24}
\end{equation}

The \( \omega \) in Equation (\ref{eq23}) is the result derived using the first-order WKB method. For the related programs of higher-order WKB methods, one can refer to the work of Konoplya et al.\cite{Konoplya:2019hlu}, which provides a detailed description of the application of higher-order WKB formulas and Padé approximation techniques. Equation (\ref{eq19}) is used to analyze physical systems governed by the Schrödinger-like equation and quasinormal boundary conditions, playing a key role particularly in studying quantum resonance phenomena at the top of one-dimensional potential barriers. In the following sections, the key numerical results obtained through the WKB method and their physical significance will be presented.

\subsection{\label{sec:level4.2} Prony method}

The Prony method\cite{Dubinsky:2024gwo}\cite{Berti:2007dg} is a dynamic signal analysis tool that precisely extracts features such as frequency and decay rates by decomposing complex signals into a superposition of exponentially decaying sinusoidal waves. It is particularly suitable for nonlinear decaying signals and has been successfully applied to the analysis of black hole QNMs frequencies\cite{Liu:2024xcd}\cite{Liu:2023vno}\cite{Yang:2022ifo}. The Prony method does not require initial parameter guesses and is computationally efficient, but it is sensitive to noise. In high-noise environments, it is often combined with methods such as the matrix pencil method or singular value decomposition to improve accuracy. Reference\cite{Gundlach:1993tp} proposes an efficient time-domain integration method for analyzing the master wave equation (\ref{eq9}). By introducing the null coordinates \( du = dt - dr_* \) and \( dv = dt + dr_* \), Equation (\ref{eq9}) can be reformulated in the following form:
\begin{equation}
\left( 4 \frac{\partial^2}{\partial u \partial v} + V(u, v) \right) \varphi(u, v) = 0.
	\label{eq25}
\end{equation}
Using the finite difference scheme, the data at position \( N \) can be derived from the data at positions \( W \), \( E \), and \( S \), thereby expressing the above equation in the following form:
\begin{equation}
		\varphi(N) = \varphi(W) + \varphi(E) - \varphi(S) - \frac{h^2}{8} V(S) \left( \varphi(W) + \varphi(E) \right).
		\label{eq26}
\end{equation}
In the above formula, the coordinates of the points are defined as \( S = (u, v) \), \( W = (u + h, v) \), \( E = (u, v + h) \), and \( N = (u + h, v + h) \). Using the integration method of Equation (\ref{eq26}), the value of the wave function \( \varphi \) can be calculated based on the region defined by the initial conditions \( u = u_0 \) and \( v = v_0 \). Through this calculation process, the properties of the QNMs can be further analyzed, and their frequencies can be extracted. For this purpose, the late-time signal at a specific \( r_* \) can be expressed as a superposition of multiple QNMs:
\begin{equation}
	\varphi(t) =\sum_{j=1}^{p} C_j e^{-i \omega_j t}.
	\label{eq27}
\end{equation}
Here, \( C_j \) and \( \omega_j \) represent the amplitude and complex frequency of the mode, respectively. To facilitate the numerical computation and processing of the signal, the time \( t \) is converted into a discrete form \( t = nh \) (where \( h \) is the time step and \( n \) is an integer). This operation allows the continuous signal to be expressed as a discrete sequence, enabling better utilization of numerical methods for operations and supporting the storage and analysis of the signal. On this basis, the form of the discrete signal \( x_n \) can be obtained:
\begin{equation}
	x_n = \varphi(nh) = \sum_{j=1}^{p} C_j e^{-i \omega_j nh} = \sum_{j=1}^{p} C_j z_j^n .
	\label{eq28}
\end{equation}
Next, we use the known \( x_n \) to construct a polynomial \( A(z) \) such that the modal roots \( z_j \) of the signal become its roots. Through these roots, the corresponding complex frequencies \( \omega_j \) can be further calculated. To achieve this, the polynomial function \( A(z) \) can be defined in the following form:
\begin{equation}
	A(z) = \prod_{j=1}^{p} (z - z_j) = \sum_{m=0}^{p} a_m z^{p-m}, \quad a_0 = 1 .
	\label{eq29}
\end{equation}
Thus, the signal satisfies
\begin{equation}
	\sum_{m=0}^{p} a_m x_{n-m} = \sum_{m=0}^{p} a_m \sum_{j=1}^{p} C_j z_j^{n-m} = \sum_{j=1}^{p} C_j z_j^{n-p} A(z_j) = 0 .
	\label{eq30}
\end{equation}
From this, we can obtain
\begin{equation}
	\sum_{m=1}^{p} a_m x_{n-m} = -x_n .
	\label{eq31}
\end{equation}
By determining the coefficients \( a_m \) of the polynomial \( A(z) \), the roots \( z_j \) of the equation \( A(z) = 0 \) can be solved. Subsequently, the frequencies of each mode can be further calculated using the following formula:
\begin{equation}
\omega_j = \frac{i}{h} \ln(z_j).
	\label{eq32}
\end{equation}
The Prony method has demonstrated excellent applicability in the study of black hole QNMs, particularly playing a crucial role in the testing of the SCCC. By accurately capturing the decay behavior of perturbation signals, this method enables researchers to efficiently extract the frequencies of QNMs, providing significant assistance in gaining deeper insights into the internal structure and dynamical evolution of black holes. The following sections will present the relevant computational results in detail.

\subsection{\label{sec:level4.3}Weak Gravity Conjecture}
In 2007, Arkani-Hamed et al. published a groundbreaking paper on the Weak Gravity Conjecture (WGC)\cite{Arkani-Hamed:2006emk}. In recent years, research on the WGC has garnered increasing attention\cite{Harlow:2022ich}, with related discussions and explorations deepening continually. The WGC is a theoretical hypothesis about the effects of quantum gravity, which asserts that the strength of gravitational interactions should not exceed that of other fundamental interactions, expressed as \( F_{\text{gravity}} \leq F_{\text{any}} \). By imposing constraints on specific physical models, the WGC provides an important approach to understanding the effects of quantum gravity in the low-energy regime. Furthermore, the WGC requires that, in a \( U(1) \) gauge theory, there exists at least one object that satisfies the following condition:
\begin{equation}
\frac{|q|}{m} \geq \left. \frac{|Q|}{M} \right|_{\text{ext}}	.
	\label{eq33}
\end{equation}
In the above expression, \( q \) represents the charge of the object, \( m \) represents its mass, and \( \left. \frac{|Q|}{M} \right|_{\text{ext}} \) denotes the charge-to-mass ratio of an extremal black hole. This relation suggests that if an object's charge-to-mass ratio is smaller than this extremal limit, it may lead to unstable physical phenomena, thereby affecting the existence of black holes.

The WGC ensures that gravity remains sufficiently strong in physical theories by comparing the strength of gauge forces and gravity. This simple yet profound idea has had far-reaching impacts across multiple areas of modern fundamental physics, including cosmology, particle physics, black holes, and scattering amplitudes. In the study of the  SCCC, the WGC plays a critical role. Specifically, when considering black holes in extremal states, the WGC ensures compliance with the SCCC by constraining the charge-to-mass ratio of charged and massive scalar fields. More concretely, the WGC offers a new perspective for exploring the SCCC and provides conditions under which extremal black holes adhere to the conjecture. This novel approach opens up new directions for investigating the validity and applicability of the SCCC in different physical scenarios.

Although the WGC has triggered extensive research and led to variants such as the ``tower WGC''\cite{Montero:2016tif} and ``sublattice WGC,''\cite{Heidenreich:2016aqi}\cite{Heidenreich:2019zkl} each proposing different constraints in various physical contexts, no version has yet been rigorously proven mathematically. Existing studies, while supportive of certain versions, still face challenges such as the lack of precise \( O(1) \) factors\cite{Cheung:2014ega}\cite{Andriolo:2020lul} or reliance on unverified assumptions. Therefore, while the WGC has profound significance in theoretical physics, it remains a conjecture requiring further validation and development. The specific computational results will be presented in detail in the following sections.

\section{\label{sec:level5}Results Analysis}

This paper, combining numerical and analytical methods, verifies whether the Reissner-Nordström (RN) black hole corrected by a Yang-Mills field satisfies the SCCC under specific conditions. The results of the numerical analysis show that even though the RN black hole violates the SCCC in extreme cases, the RN black hole modified by the Yang-Mills field can still comply with the SCCC. This phenomenon suggests that the involvement of the Yang-Mills field may alter the properties of the black hole, making it satisfy the SCCC. Furthermore, based on an analysis grounded in the WGC, it was found that even in extreme cases of the Yang-Mills field modified RN black hole, the SCCC can still hold. This phenomenon demonstrates that under specific conditions, the SCCC remains valid for extreme black holes. In conclusion, this study not only provides a new perspective for the research on black hole theory and extreme gravitational fields but also lays an important foundation for future studies on black hole dynamics and related cosmic phenomena.

\subsection{\label{sec:level5.1}Results for neutral and massless scalar fields}

This chapter examines the applicability and validity of the SCCC by analyzing the perturbative behavior of the RN black hole modified by a Yang-Mills field. A neutral and massless scalar field is introduced in the study, and the WKB method and Prony method are employed to compute the QNMs frequencies, evaluating the dynamical stability of the black hole system. Ultimately, this paper provides solid numerical support for the verification of the SCCC by exploring the perturbative characteristics of the black hole. The following table presents the perturbation decay rate and the ratio of gravitational blueshift effects, \( \beta \), under different parameters \( p \) and Yang-Mills field charge \( q_{YM} \) for the Yang-Mills field modified RN black hole. Under the assumptions of electromagnetic charge \( Q = 1 \) and mass \( M = 1 \), the data are calculated using the WKB method and the Prony method. To further investigate the physical implications of these results, we will conduct a detailed analysis and discussion based on the data in the table. Specifically, we will focus on whether these numerical results align with the expectations of the SCCC, thereby verifying the validity of the SCCC in the background of the Yang-Mills field modified RN black hole. By comparing the QNMs frequencies obtained from different methods, we can comprehensively evaluate the dynamical stability and perturbation characteristics of the black hole system. This analysis not only helps to confirm the validity of the SCCC but also reveals the potential impact of the Yang-Mills field on black hole perturbative behavior.

\begin{table}[htbp] 
	\centering
	\begin{minipage}{0.5\textwidth}
		\centering
		\captionsetup{justification=raggedright, singlelinecheck=false}
		\renewcommand{\arraystretch}{1.5} 
		\begin{tabular}{c c c c c c}
			\hline\hline
			\rule{0pt}{0pt} 
			$q_{YM}$ & Method & $l=0$ & $l=1$ & $l=10$ & $l=20$ \\
			\hline
			\multirow{2}{*}{0.0035} & WKB & 0.765552 & 0.666077 & 0.654314 & 0.653336 \\
			& Prony & 0.749611 & 0.673778 & 0.652441 & 0.637887 \\
			\hline
			\multirow{2}{*}{0.0175} & WKB & 0.396797 & 0.341531 & 0.337924 & 0.337867 \\
			& Prony & 0.389389 & 0.341475 & 0.338643 & 0.336985 \\
			\hline
			\multirow{2}{*}{0.035} & WKB & 0.355744 & 0.245574 & 0.242103 & 0.241775 \\
			& Prony & 0.360214 & 0.249235 & 0.241922 & 0.238469 \\
			\hline
		\end{tabular}
		\caption{The table shows the calculated results of \( \beta \) as a function of the Yang-Mills field charge \( q_{YM} \) in the Yang-Mills field modified RN black hole, under the conditions of electromagnetic charge \( Q = 1 \), mass \( M = 1 \), and parameter \( p = 0.3 \).}
		\label{tabI}
	\end{minipage}
\end{table}

\begin{table}[htbp] 
	\centering
	\begin{minipage}{0.5\textwidth}
		\centering
		\captionsetup{justification=raggedright, singlelinecheck=false}
		\renewcommand{\arraystretch}{1.5} 
		\begin{tabular}{c c c c c c}
			\hline\hline
			\rule{0pt}{0pt} 
			$q_{YM}$ & Method & $l=0$ & $l=1$ & $l=10$ & $l=20$ \\
			\hline
			\multirow{2}{*}{0.01} & WKB & 0.970447 & 0.917229 & 0.887955 & 0.887825 \\
			& Prony & 0.972061 & 0.916553 & 0.885905 & 0.875951 \\
			\hline
			\multirow{2}{*}{0.035} & WKB & 0.438971 & 0.414207 & 0.407728 & 0.407679 \\
			& Prony & 0.439297 & 0.417152 & 0.402269 & 0.400585 \\
			\hline
			\multirow{2}{*}{0.05} & WKB & 0.363749 & 0.321365 & 0.319129 & 0.319093 \\
			& Prony & 0.364806 & 0.328302 & 0.318281 & 0.315009 \\
			\hline
		\end{tabular}
		\caption{The table shows the calculated results of \( \beta \) as a function of the Yang-Mills field charge \( q_{YM} \) in the Yang-Mills field modified RN black hole, under the conditions of electromagnetic charge \( Q = 1 \), mass \( M = 1 \), and parameter \( p = 0.5 \).	}
		\label{tabII}
	\end{minipage}
\end{table}

\begin{table}[htbp] 
	\centering
	\begin{minipage}{0.5\textwidth}
		\centering
		\captionsetup{justification=raggedright, singlelinecheck=false}
			\renewcommand{\arraystretch}{1.5} 
		\begin{tabular}{c c c c c c}
			\hline\hline
			\rule{0pt}{0pt} 
			$q_{YM}$ & Method & $l=0$ & $l=1$ & $l=10$ & $l=20$ \\
			\hline
			\multirow{2}{*}{0.01} & WKB & 1.029482 & 0.959841 & 0.952426 & 0.942202 \\
			& Prony & 1.042304 & 0.965205 & 0.937639 & 0.931153 \\
			\hline
			\multirow{2}{*}{0.035} & WKB & 0.336248 & 0.318449 & 0.314627 & 0.314171 \\
			& Prony & 0.338007 & 0.319496 & 0.313311 & 0.312015 \\
			\hline
			\multirow{2}{*}{0.05} & WKB & 0.242004 & 0.221556 & 0.219614 & 0.219569 \\
			& Prony & 0.247037 & 0.222024 & 0.217735 & 0.216372 \\
			\hline
		\end{tabular}
		\caption{The table shows the calculated results of \( \beta \) as a function of the Yang-Mills field charge \( q_{YM} \) in the Yang-Mills field modified RN black hole, under the conditions of electromagnetic charge \( Q = 1 \), mass \( M = 1 \), and parameter \( p = 0.7\).}
		\label{tabIII}
	\end{minipage}
\end{table}

\begin{figure*}
	\centering
	\resizebox{\linewidth}{!}{\includegraphics*{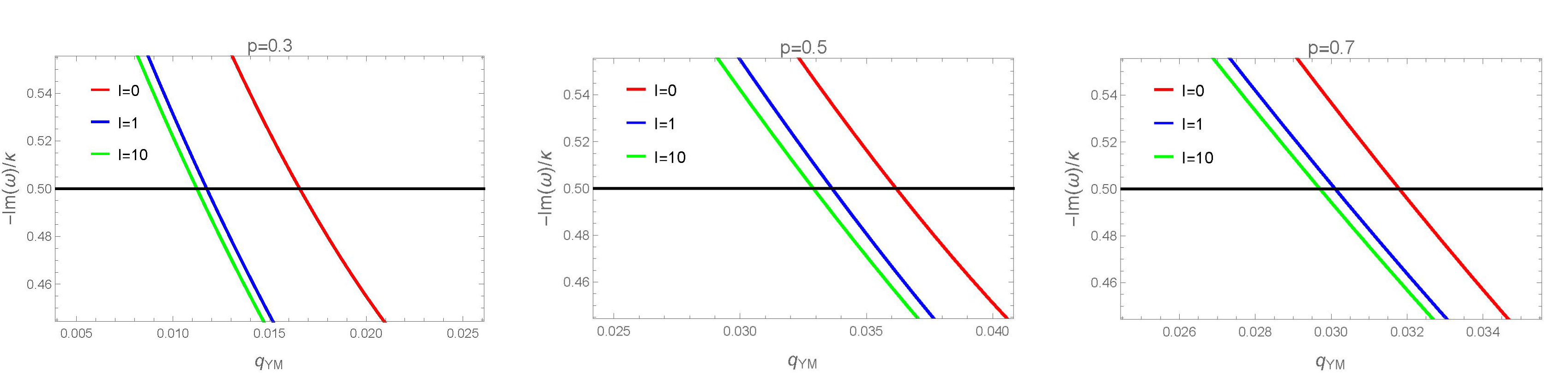}}
	\captionsetup{justification=raggedright, singlelinecheck=false}  
	\caption{The variation of $B$ with respect to the Yang-Mills field charge $q_{YM}$ for different values of the parameter $p$ in the case of the Yang-Mills field modified RN black hole with electric charge $Q=1$ and mass $M=1$.}
	\label{fig2}
\end{figure*}

From the data in Tables  \text{\ref{tabI}}, \text{\ref{tabII}} and \text{\ref{tabIII}}, and FIG \ref{fig2}, it can be observed that in the Yang-Mills field modified RN black hole, under the conditions of black hole mass \( M = 1 \) and electromagnetic charge \( Q = 1 \), and with the parameter \( p \) fixed, as the Yang-Mills field charge \( q_{YM} \) gradually increases, the RN black hole transitions from initially violating the SCCC to eventually complying with it. More importantly, as \( q_{YM} \) further increases, the likelihood of the extreme RN black hole adhering to the SCCC increases significantly. This result indicates that the Yang-Mills field introduces new spacetime geometric properties to the classical RN black hole. It not only alters the relative position of the event horizon and the Cauchy horizon but also significantly affects the behavior of scalar field perturbations on the RN black hole and its compliance with the SCCC. Combined with the results in Figure \ref{fig1}, it can be seen that the impact of the Yang-Mills field on the RN black hole under extreme conditions is particularly significant. Even when the RN black hole is in an extreme state, as the Yang-Mills field charge \( q_{YM} \) increases, the black hole's state can gradually stabilize, ultimately making it satisfy the requirements of the SCCC. This finding reveals the profound influence of the Yang-Mills field on the horizon structure of black holes, providing a concrete example of how an RN black hole can transition from an extreme state to a stable state. It also offers critical theoretical support for the completeness and consistency of general relativity.

\subsection{\label{sec:level5.2}Results for charged massive scalar fields}

This section explores the perturbation characteristics of a charged and massive scalar field near the Yang-Mills field modified RN black hole, aiming to evaluate its impact on the validity of the SCCC. Considering the fine-structure constant \( e^2 / c\hbar \simeq 1/137 \) and the fact that the charge \( Q \) of the black hole is usually very large, for a scalar field with a slight charge, its charge \( q \) would satisfy the condition \( qQ \gg 1 \). To avoid the Schwinger effect\cite{Hod:1999nb} imposing constraints on the black hole's electric field, the electric field strength of the black hole must be much lower than the critical value for Schwinger discharge, requiring the condition \( Q / r_{h_+}^2 \ll m^2 / q \), where \( r_{h_+} \) represents the radius of the event horizon, and \( m \) and \( q \) are the mass and charge of the scalar field, respectively. Thus, the constraint relationship \( m^2 r_{h_+}^2 \gg qQ \gg 1 \) can be derived. Based on the research by Jafar Saeed Noori Gashti et al.\cite{Sadeghi:2023lkz}, the following constraint conditions can be introduced to simplify the analysis of the QNM perturbations of the charged, massive scalar field near the Yang-Mills field modified RN black hole:
\begin{equation}
	m^2 r_{h_+}^2 \gg l(l+1); \quad m^2 r_{h_+}^2 \gg 2k_{r_{h_+}} r_{h_+}.
	\label{eq34}
\end{equation}
By introducing two approximation conditions, Equation (\ref{eq34}) simplifies the effective potential (\ref{eq13}) of the charged and massive scalar field, thereby significantly reducing the complexity of calculating QNMs. 

To study the linear dynamical properties of the charged and massive scalar field near the horizon of the Yang-Mills field modified RN black hole, the imaginary part of its QNMs needs to be calculated. First, the perturbative characteristics of the region are analyzed using the radial potential (\ref{eq13}), and the electric potential in the region (\ref{eq34}) is treated as the effective potential. Then, the imaginary part of the perturbative QNMs near the black hole horizon is calculated using the WKB method. Assuming that the Yang-Mills field modified RN black hole has a maximum effective potential at the point \( r_0 \) near the event horizon, the position of this maximum effective potential can be determined using Equation (\ref{eq13}) and the condition \( V'(r_0) = 0 \)\cite{Sadeghi:2023lkz}:
\begin{equation}
	r_0 = \frac{q^2 Q^2}{qQ\omega - m^2 r_{h_+}^2 k_{r_{h_+}}}.
	\label{eq35}
\end{equation}
Using Equations (\ref{eq13}), (\ref{eq20}), (\ref{eq21}), (\ref{eq22}), (\ref{eq34}), and (\ref{eq35})., the corresponding results can be derived:
\begin{equation}
	K \simeq \frac{k_{r_{h_+}}^2 m^4 r_{h_+}^4 qQ}{2 f_{r_0} \left( k_{r_{h_+}} m^2 r_{h_+}^2 - qQ \omega \right)^2},
	\label{eq36}
\end{equation}

\begin{widetext} 
	\begin{equation}\label{eq37}
		\begin{aligned}
			\Lambda(n) \simeq \frac{k_{r_{h_+}}^2 m^4 \left[ 17 - 60 \left( n + \frac{1}{2} \right)^2 \right] r_{h_+}^4 + 2 k_{r_{h_+}} m^2 \left[ 36 \left( n + \frac{1}{2} \right)^2 - 7 \right] q Q r_{h_+}^2 \omega f_{r_0}}{16 q Q \left( qQ \omega - 3 k_{r_{h_+}} m^2 r_{h_+}^2 \right)^2},
		\end{aligned}
	\end{equation}
\end{widetext}

\begin{widetext} 
	\begin{equation}\label{eq38}
		\begin{aligned}
			\Omega(n) \simeq \frac{15 k_{r_{h_+}}^4 m^8 \left[ 148 \left( n + \frac{1}{2} \right) - 41 \right] r_{h_+}^8 + 12 k_{r_{h_+}}^3 m^{6} \left[ 121 - 420 \left( n + \frac{1}{2} \right)^2 \right] q Q r_{h_+}^6 \omega  (- \left( n + \frac{1}{2} \right) Q^3 q^3 f_{r_0}^2)}{64 q^5 Q^5 \left( k_{r_{h_+}} m^2 r_{h_+}^2 - qQ \omega \right)^4}.
		\end{aligned} 
	\end{equation}
\end{widetext}
Subsequently, it is necessary to derive the specific form of the QNMs in the system to further explore the validity of the SCCC. For this purpose, \(\text{Im}(\omega)\) can be calculated by combining Equations(\ref{eq19}), (\ref{eq36}), (\ref{eq37}), and (\ref{eq38}).
\begin{equation}
	\begin{aligned}
		\omega \simeq \frac{qQ}{r_{h_+}} - \frac{2k_{r_{h_+}} m^2 r_{h_+}^2}{qQ} 
		\left[ 1 - \frac{14400}{11644} 
		\left( \frac{n + \frac{1}{2}}{qQ} f_{r_0} \right)^4 \right] \\
		- i \left[ 4 f_{r_0} k_{r_{h_+}} 
		\left( n + \frac{1}{2} \right) \frac{m^2 r_{h_+}^2}{q^2 Q^2} 
		\left( 1 - \frac{34qQ f_{r_0}^4}{11644} \right) \right].
	\end{aligned}
	\label{eq39}
\end{equation}
Thus, the relationship between the imaginary part of the QNMs and the surface gravity of the event horizon can be expressed as:
\begin{equation}
	\beta = \frac{-\text{Im}(\omega)}{k_{r_{h_+}}} \simeq 2 f_{r_0} \frac{m^2 r_{h_+}^2}{q^2 Q^2} 
	\left[ 1 - \frac{34 qQ f_{r_0}^4}{11664} \right].
	\label{eq40}
\end{equation}
Since \( r_0 \) is very close to the event horizon \( r_{h_+} \), it follows that \( f_{r_0} \ll 1 \). When \( \beta < 1/2 \), the SCCC will hold, and the condition \( \frac{m^2 r_{h_+}^2}{q^2 Q^2} < 1 \) further ensures the validity of the SCCC. Therefore, the condition for the SCCC to hold can be derived from Equation (\ref{eq40}) as\cite{Sadeghi:2023lkz}:
\begin{equation}
	\frac{q}{m} \geq \frac{r_{h_+}}{Q}.
	\label{eq41}
\end{equation}
If in Equation (\ref{eq40}), \( qQ \) satisfies \( qQ < 2\sqrt{f_{r_0}} \, m r_{h_+} \), it is possible for \( \beta > 1/2 \) to occur, which would lead to the SCCC no longer being valid. Next, we will refer to the previously derived method to study the metric equation (\ref{eq5}) in this paper. When the metric function (\ref{eq5}) of the black hole satisfies the relations \( f(r) = 0 \) and \( f'(r) = 0 \), the extremal expressions for the mass \( M \) and electromagnetic charge \( Q \) of the Yang-Mills field modified RN black hole can be obtained as:
\begin{equation}
	M = \frac{-2 Q_{YM} \, p \, r^3 + 2 Q_{YM} \, r^3 + r^{4p + 1}}{r^{4p}}.
	\label{eq42}
\end{equation}

\begin{equation}
	Q = \sqrt{\frac{-4 Q_{YM} \, p \, r^4 + 3 Q_{YM} \, r^4 + r^{4p + 2}}{r^{4p}}}.
 \label{eq43}
\end{equation}
Here, we choose the parameter \( p = \frac{1}{2} \) because it satisfies the energy and causality conditions in general relativity, significantly alters the spacetime structure of the original RN black hole, and ensures the related problems remain tractable. Compared to other values of \( p \), the choice \( p = \frac{1}{2} \) does not lead to overly complex or difficult-to-understand solutions. Instead, it provides a concise and meaningful analytical solution, particularly when describing the event horizon radius \( r_{h_+} \) and the Cauchy horizon radius \( r_{h_-} \). From equation (\ref{eq5}), we can obtain the expressions for these two horizon radii as:
\begin{equation}
	r_{h_\pm} = \frac{M \pm \sqrt{M^2 - Q^2 (1 + Q_{YM})}}{1 + Q_{YM}}
	 \label{eq44}
\end{equation}
In (\ref{eq44}), to avoid the appearance of a naked singularity and to establish a clear relationship between the two horizons, the following conditions must be satisfied: \( Q_{YM} > -1 \) and \( 0 < Q/M < \sqrt{\frac{1}{1 + Q_{YM}}} \) \cite{Gomez:2023qyv}. Therefore, from equations (\ref{eq43}) and (\ref{eq44}), the expression for the total charge \( Q_{\text{all}} \) of the Yang-Mills field modified RN black hole in the extremal case can be derived as:
\begin{equation}
	Q_{\text{all-exe}}^2 = Q^2 (1 + Q_{YM}) = Q^2 \left(1 - \frac{\sqrt{2}}{2} q_{YM}\right) = r^2 \left(1 - \frac{\sqrt{2}}{2} q_{YM}\right)^2
	\label{eq45}
\end{equation}
Substituting the obtained extremal total charge expression \( Q_{\text{all-exe}} \) (\ref{eq45}) into equation (\ref{eq40}), we can further obtain:
\begin{equation}
	\beta \simeq 2 f_{r_0} \frac{m^2}{q^2} \left(1 - \frac{\sqrt{2}}{2} q_{YM}\right)^{-2} 
	\left[1 - \frac{34 q f_{r_0}^4 r_{h_+}}{11664} \left(1 - \frac{\sqrt{2}}{2} q_{YM}\right)\right]
	\label{eq46}
\end{equation}
According to the WGC, we can obtain the condition under which the Yang-Mills field modified RN black hole respects the SCCC as:
\begin{equation}
\frac{q}{m} > \left( 1 - \frac{\sqrt{2}}{2} q_{YM} \right)^{-1}
	\label{eq47}
\end{equation}
Therefore, when \( \frac{q}{m} \) satisfies equation (\ref{eq47}), then \( \beta < \frac{1}{2} \), and the SCCC will certainly be respected. When \( \frac{q}{m} \) satisfies:

\begin{equation}
	2\sqrt{f_{r_0}} \left( 1 - \frac{\sqrt{2}}{2} q_{YM} \right)^{-1} > \frac{q}{m}.
	\label{eq48}
\end{equation}
then \( \beta > \frac{1}{2} \), and the SCCC is violated.

In extremal black holes, since the event horizon and the Cauchy horizon nearly coincide and the surface gravity approaches zero, the perturbation of neutral and massless scalar fields cannot effectively accumulate energy through the gravitational blueshift effect. Its dynamical behavior is dominated by decay effects, ultimately causing the perturbation to decay rapidly, thereby maintaining the stability of the Cauchy horizon and violating the SCCC. In contrast, the dynamics of charged and massive scalar fields are more complex. Under the conditions satisfying the WGC, although the weakened surface gravity limits the strength of the gravitational blueshift effect and the damping effect in the initial stage may suppress the amplitude growth, a large charge-to-mass ratio \( \frac{q}{m} \) allows the scalar field to accumulate energy near the Cauchy horizon through electromagnetic coupling. When the local energy density rapidly grows due to the coupling effect and exceeds a critical value, it may trigger a divergence in spacetime curvature, thereby destroying the Cauchy horizon and supporting the validity of the SCCC. It should be noted that this result strongly depends on parameters such as the black hole's mass, charge, and the initial conditions of the perturbation. The specific mechanism still requires further verification through numerical simulations or theoretical analysis.

\section{\label{sec:level6}Summary}
This paper takes the Yang-Mills modified RN black hole as the research subject, systematically analyzing the dynamical behavior of the Yang-Mills field modified RN black hole under scalar field perturbations, as well as the verification of whether the SCCC holds in specific cases. By introducing a non-Abelian gauge field, the Yang-Mills field not only alters the geometric properties of the RN black hole but also introduces two key control parameters: the coupling parameter $p$ and the Yang-Mills field charge $q_{YM}$. These parameters significantly affect the structure of the black hole horizon and its response to perturbations. The main goal of this study is to analyze whether, under the adjustment of these parameters, the Yang-Mills field modified RN black hole satisfies the SCCC, particularly when the original RN black hole is in an extreme state, exploring its potential contributions to singularity theory and gravitational dynamics.

To achieve this goal, the perturbation equations for the Yang-Mills field modified RN black hole due to a neutral, massless scalar field and a charged, massive scalar field were first derived. These equations describe the propagation characteristics of the scalar field near the black hole horizon. Then, using both the WKB method and the Prony method, the QNMs frequencies of the perturbations were approximated through analytical and numerical analysis. Additionally, the WGC provided the theoretical framework for this study, and by constraining the charge-to-mass ratio of the charged and massive scalar field, it was further demonstrated that even under extreme conditions, the Yang-Mills field modified RN black hole can still obey the SCCC.

In the perturbation analysis of the Yang-Mills field modified RN black hole with a neutral and massless scalar field, it can be seen from  Tables \text{\ref{tabI}}, \text{\ref{tabII}} and \text{\ref{tabIII}}, and Figure \ref{fig2} that, under the conditions where the Yang-Mills field modified RN black hole has mass $M=1$, electric charge $Q=1$, and a fixed parameter $p$, the increase in the Yang-Mills field charge $q_{YM}$ significantly weakens the damping effect of the perturbations and enhances the gravitational blueshift effect. This effect allows the Yang-Mills field modified RN black hole to transition from a state that violates the SCCC to one that obeys the SCCC. Furthermore, as shown in Figure \ref{fig1}, adjusting the Yang-Mills field charge $q_{YM}$ and the parameter $p$ not only alters the structural characteristics of the black hole but also potentially stabilizes the previously extreme RN black hole. This suggests that the non-linear field modified play a key role in black hole dynamics. Specifically, the external electric field of the Yang-Mills field, by adjusting the distance between the black hole horizons and enhancing the blueshift effect, makes it easier for scalar field perturbations to disrupt the Cauchy horizon, thus ensuring the SCCC. Compared to the classical RN black hole model, the Yang-Mills field modified RN black hole can maintain stability over a wider range of parameters, further extending the conditions under which the SCCC holds.

Through the perturbation of the Yang-Mills field modified RN black hole by a charged and massive scalar field and the analysis results combined with the WGC method, it can be concluded that when the charge-to-mass ratio $q/m$ of the charged and massive scalar field satisfies $\frac{q}{m} > \left( 1 - \frac{\sqrt{2}}{2} q_{YM} \right)^{-1}
$, the Yang-Mills field modified RN black hole in the extreme state will still obey the SCCC. On the other hand, when the charge-to-mass ratio $2\sqrt{f_{r_0}} \left( 1 - \frac{\sqrt{2}}{2} q_{YM} \right)^{-1} > \frac{q}{m}
$, the Yang-Mills field modified RN black hole in the extreme state will violate the SCCC. In conclusion, when the Yang-Mills field modified RN black hole is in an extreme state, if the charge-to-mass ratio $q/m$ of the charged and massive scalar field satisfies the WGC condition, it indicates that the perturbation of the scalar field can disrupt the stability of the Cauchy horizon, thus supporting the validity of the SCCC. However, when $q/m$ does not satisfy the WGC condition, the perturbation cannot disrupt the stability of the Cauchy horizon, leading to a violation of the SCCC. This result reveals the key role of the charge-to-mass ratio $q/m$ of the charged and massive scalar field in black hole dynamics and theoretically validates the applicability of the WGC for the stability of the extreme state of the Yang-Mills field modified RN black hole. At the same time, this study further deepens the understanding of the SCCC and provides new theoretical support for verifying the applicability of General Relativity in strong gravitational fields.

Through theoretical derivation and numerical simulations, this paper verifies the conditions under which the Yang-Mills field modified RN black hole supports the SCCC and reveals how the nonlinear Yang-Mills field influences the stability of the Cauchy horizon by adjusting the black hole's structural features. The results show that the Yang-Mills field significantly alters the damping effect of scalar field perturbations and the gravitational blueshift effect, causing the RN black hole, initially in an extreme state, to transition from violating the SCCC to obeying the SCCC. This finding not only provides strong support for the theoretical completeness of General Relativity but also offers important clues for studying the singularity problem of black holes under nonlinear gravitational effects. At the same time, the results further confirm that under the condition of obeying the WGC, the SCCC can also hold for the Yang-Mills field modified RN black hole in an extreme state, revealing the profound connection between nonlinear field corrections, General Relativity, and quantum field theory. Future research can focus on exploring the impact of other types of nonlinear fields on different black hole properties, especially in extreme states, and their effects on the stability of the Cauchy horizon and the validity of the SCCC.

\vspace{2em} 
\section{\label{sec:level7}Acknowledgements}
\noindent

We acknowledge the anonymous referee for a constructive report that has significantly improved this paper.This work was  supported by Guizhou Provincial Basic Research Program(Natural Science)(Grant No.QianKeHeJiChu-[2024]Young166),  the Special Natural Science Fund of Guizhou University (Grant No.X2022133), the National Natural Science Foundation of China (Grant No.12365008) and the Guizhou Provincial Basic Research Program (Natural Science) (Grant No.QianKeHeJiChu-ZK[2024]YiBan027 and QianKeHeJiChu-ZK[2025]General Program680) .

\bibliography{rf1.bib}
\bibliographystyle{apsrev4-1}

\end{document}